%% file: main.tex
\documentclass{article}

\pdfoutput=1
\newcommand{\remove}[1]{}

\PassOptionsToPackage{square,sort,comma,numbers}{natbib}

\usepackage{DAVID_nips_2016}

\usepackage[utf8]{inputenc} 
\usepackage[T1]{fontenc}    
\usepackage{url}            
\usepackage{booktabs}       
\usepackage{amsfonts}       
\usepackage{nicefrac}       
\usepackage{microtype}      
\usepackage{subcaption}
\usepackage{textcomp}

\usepackage{amsmath}

\usepackage[pdftex]{graphicx}

\title{A Multi-Pass Approach to Large-Scale Connectomics} 

\usepackage{amssymb,tikz}  

\newboolean{showcomments}
\setboolean{showcomments}{true}
\ifthenelse{\boolean{showcomments}}
{ \newcommand{\mynote}[3]{
    \fbox{\bfseries\sffamily\scriptsize#1}
    {\small$\blacktriangleright$\textsf{\emph{\color{#3}{#2}}}$\blacktriangleleft$}}}
{ \newcommand{\mynote}[3]{}}

\author{
  AUTHORS\\
  Massachusetts Institute of Technology\\
  Computer Science and Artificial Intelligence Laboratory\\
  \texttt{email} \\
}

\begin{document}

\maketitle
 
\input{abstract_new} 
\input{introduction_new}

\input{pipeline_new}
\input{cnn_new}
\input{blockMerge_new}
\input{skeletonization_new}

\input{errorDetection}

\input{conclusions_new}


\small
\bibliography{reconstruction}
\bibliographystyle{abbrv}

\end{document}

%% file: abstract_new.tex
\begin{abstract}
The field of connectomics faces unprecedented ``big data'' challenges.  To reconstruct neuronal connectivity, automated pixel-level segmentation is required for petabytes of streaming electron microscopy data. Existing algorithms provide relatively good accuracy but are unacceptably slow, and would require years to extract connectivity graphs from even a single cubic millimeter of neural tissue. Here we present a viable real-time solution, a multi-pass pipeline optimized for shared-memory multicore systems, capable of processing data at near the terabyte-per-hour pace of multi-beam electron microscopes. The pipeline makes an initial fast-pass over the data, and then makes a second slow-pass to iteratively correct errors in the output of the fast-pass. We demonstrate the accuracy of a sparse slow-pass reconstruction algorithm and suggest new methods for detecting morphological errors. Our fast-pass approach provided many algorithmic challenges, including the design and implementation of novel shallow convolutional neural nets and the parallelization of watershed and object-merging techniques. We use it to reconstruct, from image stack to skeletons, the full dataset of Kasthuri et al.~(463 GB capturing 120,000 cubic microns) in a matter of hours on a single multicore machine rather than the weeks it has taken in the past on much larger distributed systems. \end{abstract}

%% file: introduction_new.tex
\section{Introduction} 



The field of connectomics aims to map neural circuitry at the level of synaptic connections, in order to gain new insight into the structure and emergent properties of neural computation. Connectomics began in the 1970s with an eight-year-long study of the 302 neuron nervous system of a worm~\cite{emmons2015beginning,White1}. Although this and other manual mapping efforts have led to breakthrough biological discoveries (e.g.,~\cite{chalasani2007dissecting,chen2006wiring,gray2005circuit,wen2012proprioceptive}), modern connectomics bears little resemblance to this work. In place of laborious manual tracing, cutting-edge machine learning and image-processing techniques are being developed to extract connectivity graphs automatically from nanometer-scale electronic microscopy (EM) images. This automation will allow for the high-resolution investigation of neural computation at an unprecedented scale.

Recent progress in sample preparation and multi-beam EM imaging technology has made it feasible to collect nanometer resolution imagery of a cubic millimeter of tissue, enough to contain a cortical column. At a rate of about half a terabyte an hour, this would yield a two-petabyte stack of 33,000 images within six months \cite{Microscope,Seung}.  However, the machine learning algorithms required to process such images remain a bottleneck.  Segmenting and reconstructing the 3-D neuronal structure within this cubic millimeter volume using today's fastest automated pipelines \cite{PlazaBerg2016,roncal2015automated} would take decades.  To date, the connectomics literature has largely used segmentation accuracy as the measure of performance \cite{knowles2016rhoananet, roncal2015automated}.  While it is clearly essential for algorithms to have high accuracy, it is also imperative to seek algorithms that scale to large volumes of brain tissue.

In 2014, several of the authors presented the ``big data'' challenges of connectomics and set an agenda for addressing them~\cite{Lichtman-Pfister-Shavit-Survey}. Foremost among these issues is that the petabyte scale and terabyte-an-hour pace of data that modern electron microscopes generate make local storage unlikely and transmission to an off-site large scale distributed computational facility hard, even with a dedicated fiber-optic line. Even if possible, such mass storage and compute capabilities are out of the reach of most microscopy labs in the world. So instead of a large scale distributed solution, we chose to focus our efforts on developing a pipeline to run on a shared-memory multicore system, plausibly physically located in the same room as the microscope~\cite{matveev2016}. In such a system, it is vital that reconstruction executes in real time on the continuous stream EM data. Developing a fast reconstruction pipeline is also fundamental to achieving accurate reconstruction. If on today's fastest connectomics pipeline \cite{PlazaBerg2016}, processing a terabyte requires 140 hours on 500 cores, there is no reasonable way to engage in the repeated design and testing of segmentation algorithms on even a single terabyte -- let alone the tens of terabytes that a meaningful test volume will require. 

We take a first step in addressing this problem: a pipeline for large-scale reconstruction of neuron morphology that is both (a) on par with (or better than) previous state-of-the-art~\cite{roncal2015automated} in accuracy, yet (b) requires only a fraction of the computation time. This allows us to segment and skeletonize volumes such as the full dataset of Kasthuri \emph{et al.}~\cite{kasthuri2015saturated} (463 gigabytes capturing 120,000 cubic microns) in a matter of hours on a single multicore system rather than the weeks it has taken in the past on a farm of GPUs and CPUs \cite{kaynig2015large,roncal2015automated}.


To create a viable real-time pipeline, we introduce the paradigm of \emph{multi-pass} segmentation: a \emph{fast-pass} algorithm swiftly creates an initial segmentation which is proofread by additional \emph{slow-pass} algorithms, and areas flagged as erroneous are re-segmented more carefully, as time permits.  We describe our fast-pass algorithm in detail, extending our algorithmic approach described in \cite{matveev2016}.   We then outline a framework for recurrent improvements of neurons' morphology based on automated detection and correction of errors arising in the initial reconstruction. There are several advantages to multi-pass segmentation.  
The majority of voxels are easy to label, and are assigned quickly in the first pass of the pipeline; additional slow-pass computational resources are assigned only to regions of the image with a high probability of error.  We show two techniques for error correction. The first is a technique for detecting merge errors because they form biologically implausible geometries in what otherwise might seem like valid neuronal junctions.  The second is a machine learning based approach to extending broken axon segments, tracking them through complex paths in the segmented volume and reconstructing them into a single correct axonal branch. Both of these techniques are computationally expensive, but can be applied to perform correction throughout the volume because we have saved time during the early fast-pass segmentation process. 


%% file: pipeline_new.tex
\section{A high-throughput fast-pass pipeline}
\label{sec:pipeline}
\renewcommand*\thesubfigure{\alph{subfigure}}
\captionsetup[subfigure]{skip=3pt}
\begin{figure}
\centering
\includegraphics[width=1.\linewidth]{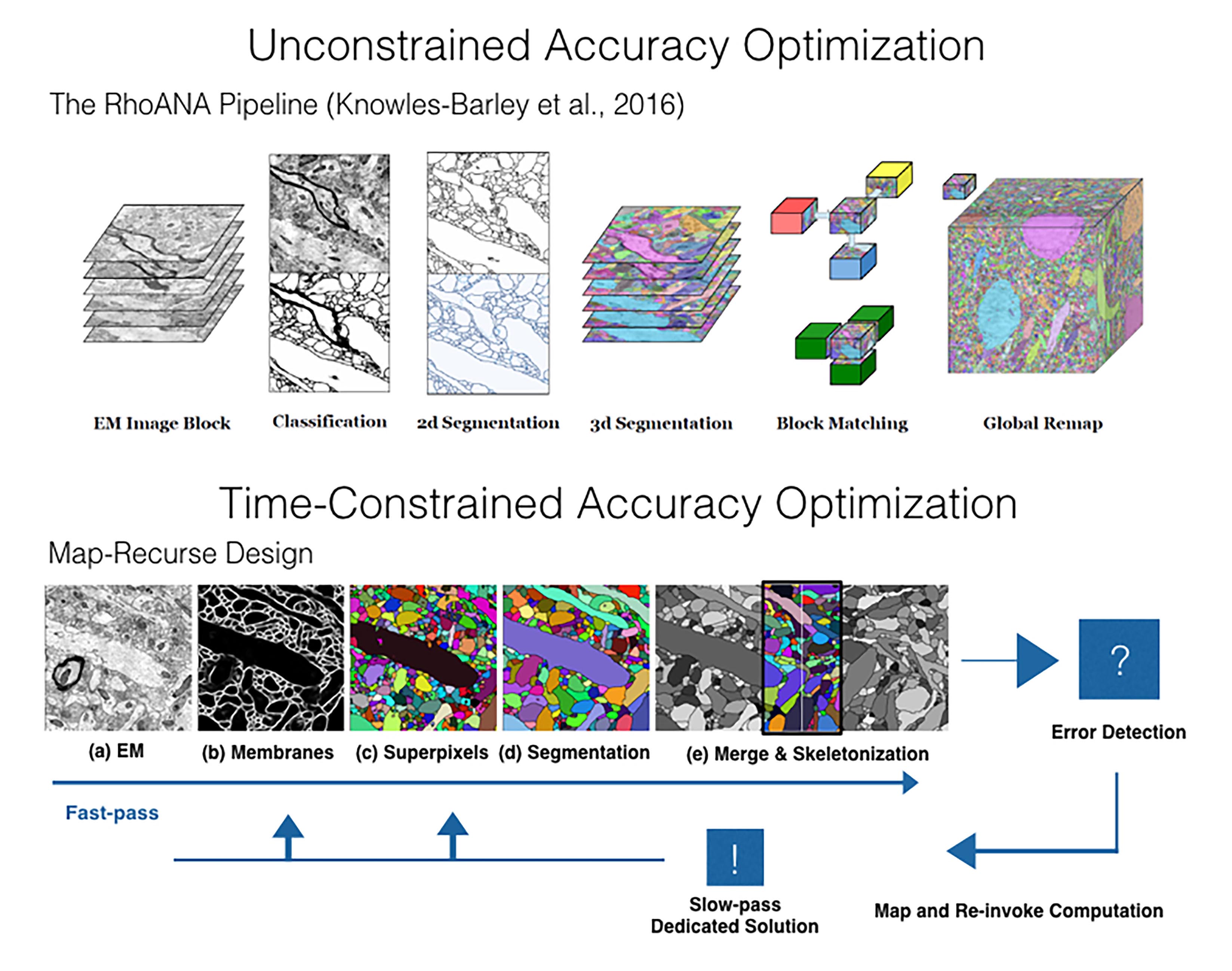}
  \caption{2-D visualization of connectomics pipeline stages. Top: RhoANA, state-of-the-art in reconstruction accuracy. A compelte system for dense neuronal morphology reconstruction running in single-pass \cite{knowles2016rhoananet}. Bottom: Our multi-pass approach to ``Connectomics-on-demand'' \cite{Lichtman-Pfister-Shavit-Survey} (a) Electron microscope (EM) image. (b) Membrane probabilities generated by convolutional neural networks. (c) Over-segmentation generated by watershed. (d) Neuron reconstruction generated by agglomeration (NeuroProof). (e) Pipeline inter-block merging: for each two adjacent blocks, the pipeline slices a boundary sub-block and executes agglomeration.\label{fig:pipeline}}
\end{figure}
\begin{figure}
\centering
\includegraphics[width=1\linewidth]{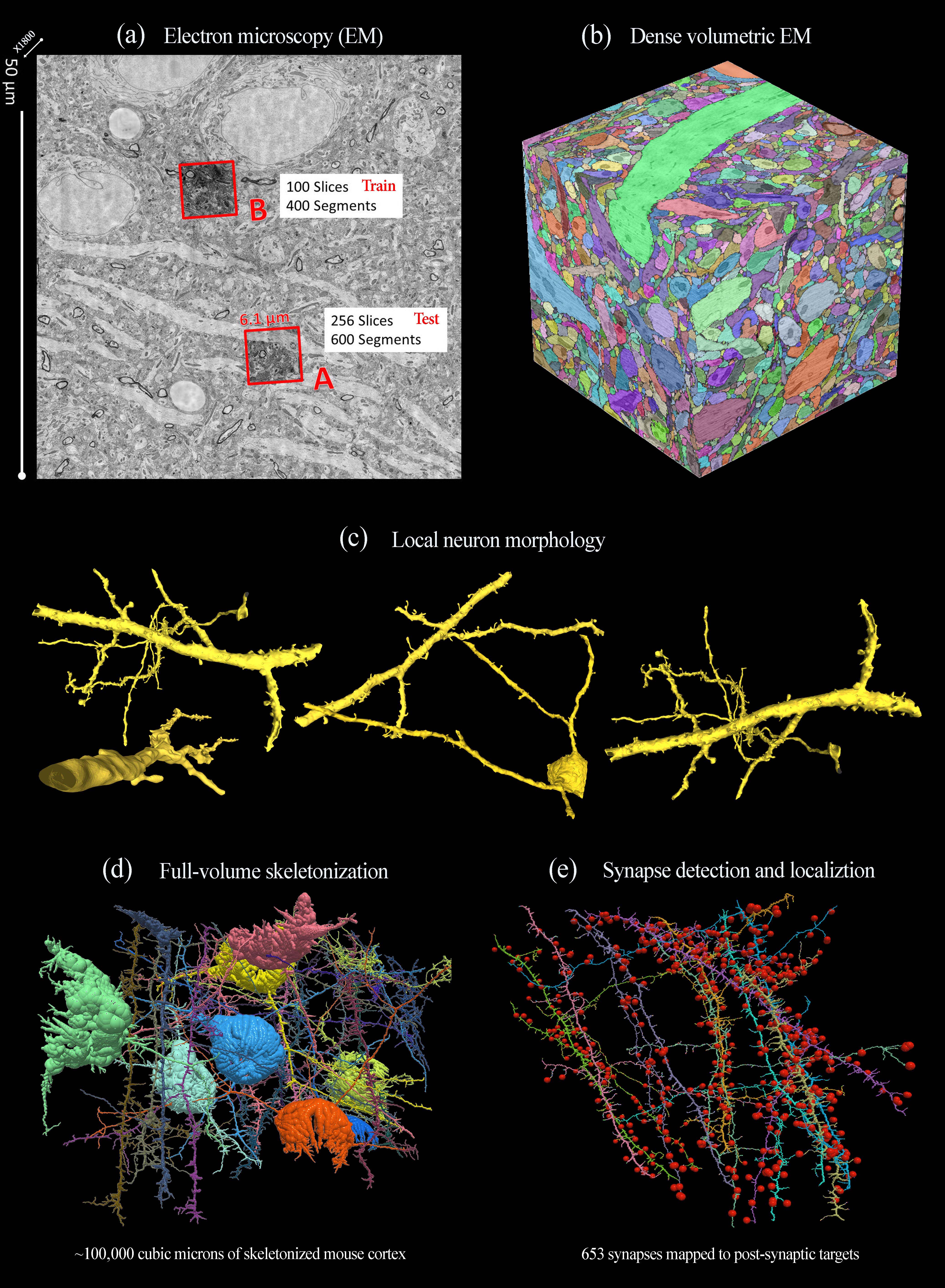}
\caption{Data Spaces Managed by our fast-pass pipeline. (a) Electron microscopy section of mouse cortex S1, Kasthuri et al. \cite{kasthuri2015saturated}. The location of the train (A. AC4) and test (B. AC3) sets are marked by red squares.  (b) Dense segmentation in which every pixel in Cartesian space is labeled in correspondance to a unique object or to extracellular space. (c). 3-D Surfaces for all reconstructed objects in S1 were computed, of which few representative were selected for visual purposes. The fast-path pipeline constructs many of the fine details of neurites' morphology to the level of spines. (d) All segmented objects were skeletonized, of which hundred large skeletons are depicted. (e) The fast-path pipeline presented herein has recently been used to predict the locations of tens of thousands of synapses in S1 with pre- and post-synaptic association \cite{shibani2016cvpr}. Nine skeletons with 653 associated synapses are shown.    \label{fig:data}}
\end{figure}
All connectomics systems are comprised of both physical and computational parts. The physical component is responsible for slicing chemically prepared brain tissue and feeding these slices to an electron microscope for imaging, and is largely beyond the scope of this paper. The computational component is responsible for reconstructing three-dimensional neurons from the thousands of resultant images. Current state-of-the-art large-scale connectomics systems have been presented by Roncal \emph{et al.}~\cite{roncal2015automated} based on the RhoANA system of Kaynig \emph{et al.}~\cite{kaynig2015large}, and Plaza and Berg \cite{PlazaBerg2016}. These systems are capable of processing terabyte-scale data in the order of weeks on distributed clusters with hundreds of CPU cores, which is orders of magnitude slower than the pace of modern multi-beam electron microscopes.

A standard computational connectomics system is illustrated in Figure~\ref{fig:pipeline}. The inputs (Figure~\ref{fig:pipeline}(a)) are EM images, which are fed through a pipeline of algorithmic modules: (b) membrane classification, (c) neuron over-segmentation, (d) agglomeration and (e) block merging. This output can then be followed by skeletonization (addressed in this paper) or synapse detection (addressed elsewhere~\cite{roncal2015vesicle,roncal2015automated,shibani2016cvpr}), which are akin to identifying nodes and edges, respectively, in a directed connectivity graph. 

This section provides a high-level overview of the design considerations, challenges and optimizations for each module. Of these, the largest computational bottleneck was the CNN-based classification of neuron membranes in step (2). We defer a detailed discussion of our CNN algorithmic and performance optimizations to Section~\ref{sec:cnns} whereas details on our multi-core implementation are presented elsewhere \cite{matveev2016}. Importantly, the CPU cycles that we save can be re-invested into error detection using higher ``neuron-level" (versus pixel-level) context. Accordingly, Section~\ref{sec:skeletons} presents a new VI-based metric for evaluating morphological reconstruction accuracy and a procedure for automated detection of spurious ``X-junctions" in neuron skeletons.

\subsection{Overview} 

\subsubsection*{Membrane classification}
\label{ss:cnnpipeline}

The first and most computationally demanding stage of today's connectomics pipeline is the mapping of EM images to membrane probability maps~\cite{PlazaBerg2016}, shown in Figure~\ref{fig:pipeline}(b). These maps capture the probability that each pixel in every input EM image is associated with a neuronal cell membrane, defining putative neuronal boundaries that are used downstream for reconstructing 3-D morphology.

We use convolutional neural networks (CNNs) for membrane classification, which are computational models motivated by the arrangement of simple and complex cells in the primary visual cortex~\cite{hubel1959receptive}. The function of these cells is approximated by alternating convolutional and pooling layers, which produce state-of-the-art performance on many image classification and segmentation problems~\cite{krizhevsky2012imagenet,long2015fully}. Although alternative methods such as conditional random fields~\cite{kaynig2015large} and random forests~\cite{PlazaBerg2016} have been proposed for use in the connectomics domain, deep CNNs running on large distributed systems are unrivaled in terms of neuron reconstruction accuracy~\cite{arganda2012segmentation,lee2015recursive}. 

\subsubsection*{Over-segmentation and agglomeration}

The next stage of our pipeline involves converting membrane probability maps into a neuron-level segmentation. First we apply a 3-D implementation of the popular watershed algorithm that is optimized to run efficiently on a multi-core machine (Figure~\ref{fig:pipeline}(c)). This stage produces an over-segmentation, such that no class label straddles more than two true neurons. We then agglomerate these segments by using NeuroProof, a software package of Parag \emph{et al.}~\cite{parag2015context} (Figure~\ref{fig:pipeline}(d)). This package applies a pre-trained random forest classifier to detect which adjacent segments should be merged on the basis of the membrane probability map. Both our watershed and NeuroProof implementations were re-implemented to maximize performance on multicore CPU systems~\cite{Victorshed,matveev2016,QuanNP}.

\subsubsection*{Block merging}

The pipeline breaks the collection of large two-dimensional EM images into blocks and segments them independently. To obtain a complete segmentation, corresponding objects in adjacent blocks must be merged. After in-block segmentation, however, the contours of predicted segments on the shared face of two adjacent blocks do not match precisely in practice. Therefore, a special algorithm is necessary for merging decisions. Plaza and Berg \cite{PlazaBerg2016} describe a hand-crafted heuristic to process complex edge cases for merging segments in adjacent blocks. \cite{knowles2016rhoananet} uses the size of overlapping volumes between segments on neighboring blocks to find a stable pairwise matching between them. Our solution, described in detail in Section \ref{sec:blockMergesec}, leverages the random-forest based agglomeration that is used to segment individual blocks.

\subsubsection*{Skeletonization and 3-D reconstruction}

Valuable information on the high-level structure of neurons can be obtained by applying skeletonization algorithms, including but not limited to their width, length and direction in 3-D space. In Section~\ref{sec:skeletons}, we describe our implementation of a fully-automated \emph{thinning} algorithm that preserves object topology while maintaining the high throughput necessary for large-scale connectomics. A different skeletonization algorithm has been presented for an application in connectomics \cite{zhao2014automatic}, however to our knowledge we are first to automatically skeletonize a volume on the scale of the Kasthuri dataset. We also describe a novel extension of VI score (commonly used for evaluating 2-D neuronal segmentation) \cite{meilua2007comparing} that allows the performance of 3-D neuron skeletonization to be evaluated.

%% file: cnn_new.tex
\subsection{Membrane classification with CNNs}
\label{sec:cnns}
Our membrane classification module is formed of two major components: the abstract CNN architecture that defines the arrangement of convolution, dilation and sampling primitives; and the concrete implementation that allows this architecture to be executed on EM data to produce membrane probability maps. Overcoming the computational bottleneck introduced by this module required rethinking both components.

\subsubsection*{A lightweight network architecture}

\begin{figure}
\centering
\includegraphics[width=1\linewidth]{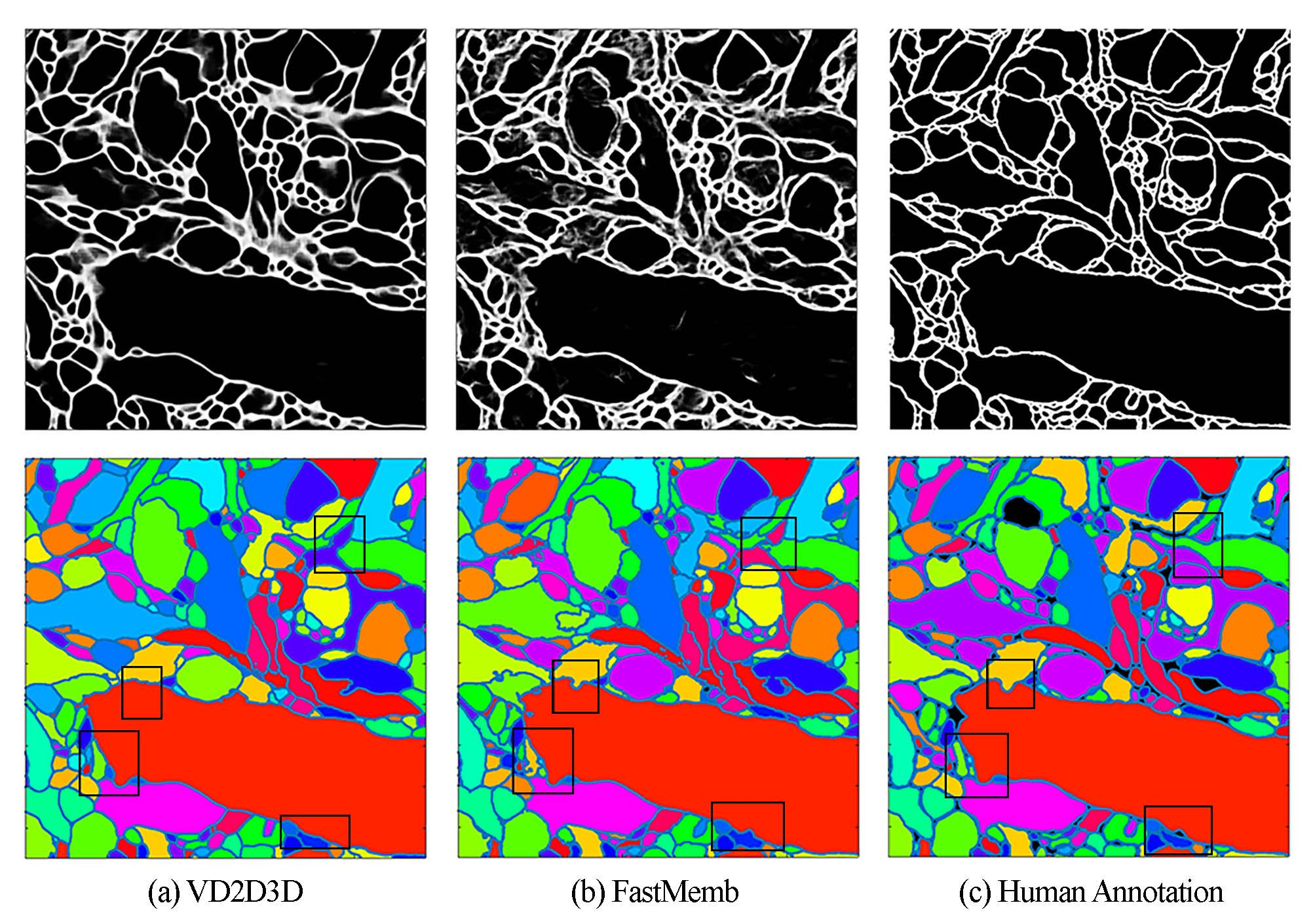}
\caption{Membrane classification and neuron segmentation using (middle) our lightweight FastMemb architecture, compared to (left) VD2D3D and (right) manually-annotated ground truth. Although the membrane predictions of VD2D3D are visually smoother, the downstream segmentation is qualitatively similar and empirically worse in terms of the VI measure (more merge errors).}
\label{fig:segmentations}
\end{figure}

Key to our performance is an architecture that runs contrary to the current trend of using increasingly deeper and more complex networks to deliver improved accuracy. This deeper-is-better trend has been central to connectomics pipeline development, as evidenced by the most accurate deep networks such as the N4 by Ciresan \emph{et al.}~\cite{ciresan2012deep}, VD2D3D by Lee~\emph{et al.}~\cite{lee2015recursive} and U-Net by Ronneberger~\emph{et al.}~\cite{ronneberger2015u}. However, we believe that CNNs producing accurate membrane classifications are not a solution unless they are efficient enough to be applied to large image stacks. Moreover, this accuracy in itself is not the whole solution because once a large volume is segmented and skeletonized, many of the errors detected are on a global scale and are not fixable by simply improving pixel-level performance \cite{Lichtman-Pfister-Shavit-Survey}.

An early version of the CNN architecture developed herein, \emph{FastMemb}, has already proved useful in augmenting the largest dense reconstruction of a piece of mammalian cortex~\cite{kasthuri2015saturated,github-knowles}. FastMemb consists of only 3 layers of alternating convolution/max-pool pairs aggregated with a max-out function \cite{goodfellow2013maxout}. Convolution layers use $4\times4$ kernels with stride 1 and 32 channels (features), which combined with stride $2\times2$ fully-convolutional max-pooling yields a $53\times53$ field of view. We note that the max-pool/max-out combination introduces sufficient non-linearity that the addition of ReLU or other transfer functions is empirically unnecessary.  Figure~\ref{fig:segmentations} provides a visual comparison of membrane prediction and downstream neuron segmentation using (middle) FastMemb compared to (left) VD2D3D and (right) manually-annotated ground truth.

\subsubsection*{Architecture evaluation}

The standard metric for evaluating neuron segmentation in connectomics literature is the variation of information (VI) score~\cite{ciresan2012deep,kaynig2015large,meilua2007comparing}. This is a measure of distance between segmentations $S_1$ and $S_2$ (\emph{i.e.}~prediction and ground truth) that we seek to minimize:
\begin{equation}
VI = \underbrace{\left(H(S_1) - I(S_1, S_2)\right)}_{VI_\mathrm{merge}} + \underbrace{\left(H(S_2) - I(S_1, S_2)\right)}_{VI_\mathrm{split}},
\label{eq:vi}
\end{equation}
where $H(\bullet)$ denotes the entropy and $I(\bullet, \bullet)$ the mutual information.  We denote the first term in the definition of $VI$ by $VI_\mathrm{merge}$ since it is measures \emph{merge errors} (incorrect merges) in $S_1$ with respect to $S_2$, whereas the second term $VI_\mathrm{split}$ measures \emph{split errors} (incorrect splits) in $S_1$ with respect to $S_2$.

We evaluate the segmentation performance of FastMemb (70k parameters) compared to previous gold-standard networks: N4 (220k parameters) by Ciresan \emph{et al.}~\cite{ciresan2012deep}, the winning architecture from the ISBI12 2D EM segmentation challenge~\cite{arganda2012segmentation}; and VD2D (230k parameters) and VD2D3D (310k parameters) by Lee~\emph{et al.}, which were recently demonstrated to yield improved segmentation performance~\cite{lee2015recursive}. We also considered the recent U-Net by Ronneberger~\emph{et al.}~\cite{ronneberger2015u}, but with two orders of magnitude more parameters than VD2D3D it currently seems infeasible to apply this network to the high-throughput connectomics domain. Each network was trained and tested on independent 6 nm-resolution volumes of the Kasthuri EM dataset~\cite{kasthuri2015saturated} known as AC4 (100 images) and AC3 (256 images) respectively. Networks were trained using the same parameters as Lee \emph{et al.}~\cite{lee2015recursive} (learning rate $= 0.01$, momentum $= 0.9$). FastMemb, N4 and VD2D3D were each trained for 90,000 iterations, with the latter recursively trained on the output of the simpler VD2D trained for 60,000 iterations as described in~\cite{lee2015recursive}.

Our performance results are presented in Table~\ref{tab:results1}. It is evident that the reconstruction accuracy produced from the lightweight FastMemb is at least as good as those of the larger N4, VD2D and VD2D3D networks, while executing 3.6x, 5.3x and 6.3x faster respectively. 

\begin{table}[h!]
\centering
\caption{Neuron segmentation performance resulting from different CNN architectures, evaluated using the VI score (smaller is better) as per previous literature~\cite{ciresan2012deep,kaynig2015large}.\\\hspace{\textwidth}}
\label{tab:results1}
\begin{tabular}{c|c|ccc}
           & Parameters & VI              & VI$_\mathrm{split}$   & VI$_\mathrm{merge}$  \\ \hline
FastMemb & 70k                 & \textbf{1.8364} & 1.3002                 & \textbf{0.5363}        \\
N4         & 220k                & 2.2024          & 1.2790                 & 0.9233                 \\
VD2D       & 230k                & 1.9292          & 1.3775                 & 0.5517                 \\
VD2D3D     & 310k                & 1.9231          & \textbf{1.2474}        & 0.6757                
\end{tabular}
\end{table}

\subsubsection*{Accelerating dense computation}

Reducing execution time by adopting a lightweight architecture is just one part of our solution. It is also necessary to implement this in a way that removes redundant computations for the overlapping fields-of-view of adjacent pixels to effectively utilize available CPU resources.

As opposed to traditional sliding-window approaches to convolution (\emph{patch calculation}), \emph{dense computation} removes redundant calculations between overlapping fields-of-view by generating all pixel-level labels in a single pass. The most common approach to enabling dense computation is to replace max-pooling layers with \emph{max-filtering}~\cite{lee2015recursive,zlateski2015znn}. More recently, Giusti~\emph{et al.} proposed a fragmentation scheme that allows for optimal dense computation in the presence of max-pooling layers (\emph{max-pooling fragments}~\cite{giusti2013fast,masci2013fast}). We generalize this scheme by decomposing all $k$-stride operations into two components: a 1-stride sliding-window computation (of the specific operation) followed by $k$-factor sub-sampling. Each operation yields $k^2$ independent submatrices, such that downstream computations (arbitrary combinations of convolution, spatial dilation and spatial sampling primitives) can be recursively computed in parallel by separate threads. Submatrices are interleaved post-softmax to produce the final dense output. In the specific case of FastMemb, dense computation requires approximately 284x fewer floating point multiplications than traditional sliding-window convolution~\cite{matveev2016}.


\subsubsection*{A subsampling CNN for 3 nm data}
\begin{figure}[t]
\centering
\includegraphics[width=0.9\linewidth]{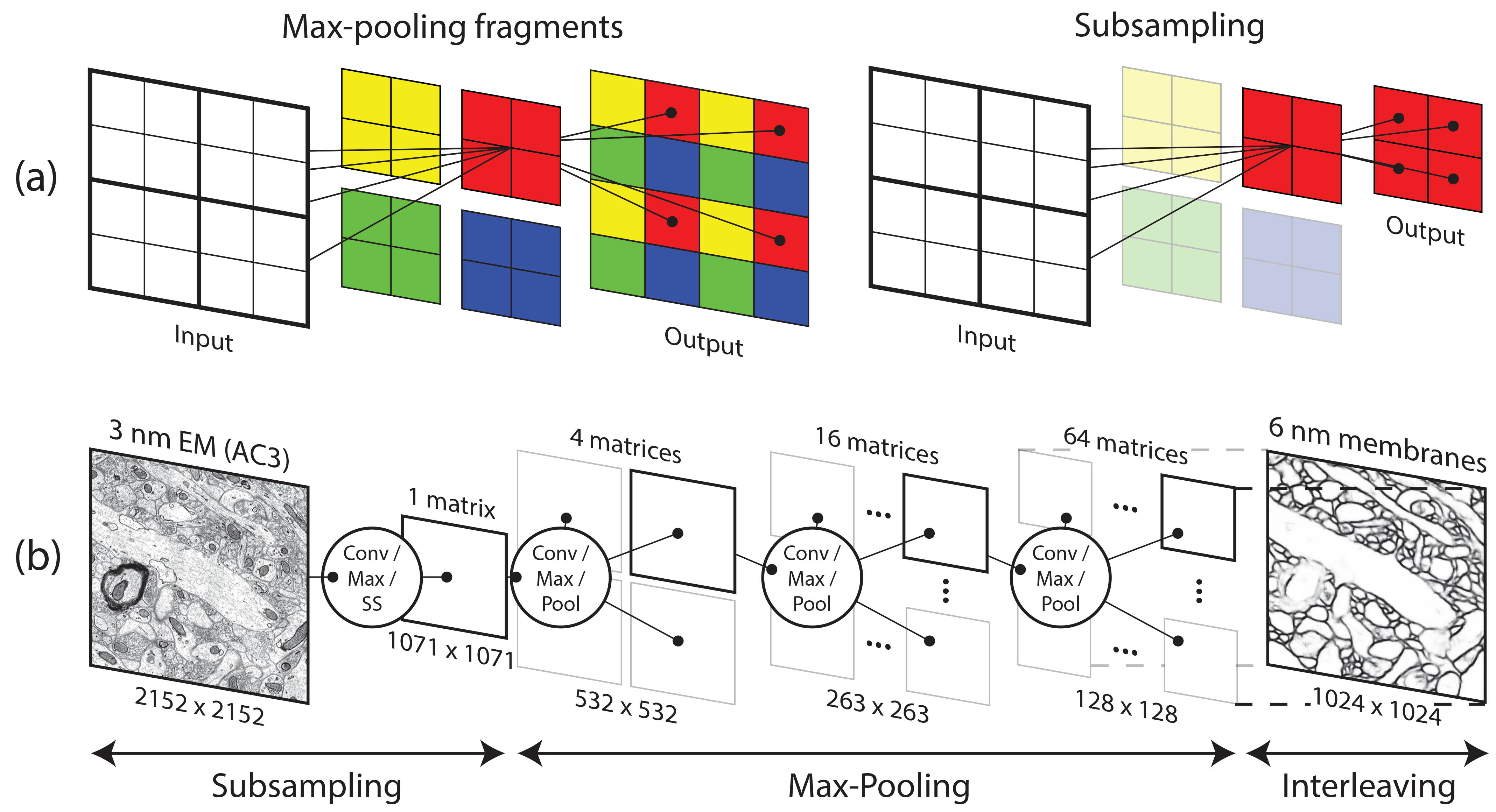}
\caption{(a) Fully convolutional max-pooling with interleaving (\emph{max-pooling fragments}~\cite{giusti2013fast,masci2013fast}) preserves input size. As intermediate matrices are calculated and stored independently, the recursion tree may be pruned to subsample the image. (b) The FastMemb$_\mathrm{ss}$ CNN architecture makes use of both subsampling and max-pooling layers to optimize speed and accuracy.}
\label{fig:cnn_merged}
\end{figure}
Our generalized approach to max-pooling fragments for dense computation can be extended to efficiently process higher resolution data. The basic approach to capturing 3 nm ($2048\times2048$) EM is to extend the FastMemb architecture (herein FastMemb$_\mathrm{6nm}$) with an additional pair of convolution ($k=4$, 32 channels) and max-pooling (2-stride) layers, aggregated by a max-out function. We call this architecture FastMemb$_\mathrm{3nm}$. Instead, we investigate the effect of replacing the top-level max-pooling layer with a subsampling layer, as shown in Figure~\ref{fig:cnn_merged}. Both variants upscale to a $105\times105$ field-of-view over the 3 nm EM (necessary to resolve equivalent membrane context), with the important difference that FastMemb$_\mathrm{ss}$ produces a sub-sampled output probability map at the same 6 nm resolution and thus requires far fewer FLOPs (17\% versus 486\% more than the baseline FastMemb$_\mathrm{6nm}$). 

The segmentation performance for the 3-and-6 nm FastMemb variants are presented in Table~\ref{tab:sampling}. As expected, presenting the CNN with higher resolution data leads to segmentation improvement. However, this is also associated with a 5.86x increase in required FLOPs (4x increase in image size plus additional conv/pool layer to scale field-of-view). By applying our dense subsampling layer, FastMemb$_\mathrm{ss}$ is able to improve baseline performance on both metrics for substantially fewer additional FLOPs (1.17x versus 5.86x). Interestingly, this approach also leads to marginal improvement over the more computationally expensive FastMemb$_\mathrm{3nm}$.

\begin{table}[h!]
\centering
\caption{Segmentation performance of FastMemb CNN for 3-versus-6 nm AC3 EM data. Introducing a subsampling layer yields better performance for substantially fewer FLOPs than naive 3 nm integration.\\\hspace{\textwidth}}
\label{tab:sampling}
\begin{tabular}{c|c|ccc}
                & FLOPs          & VI              & VI$_\mathrm{split}$       & VI$_\mathrm{merge}$       \\ \hline
FastMemb$_\mathrm{6nm}$ & baseline       & 1.8364          & 1.3002          & 0.5363          \\
FastMemb$_\mathrm{3nm}$ & 5.68x          & 1.6942          & 1.2327          & 0.4615          \\
FastMemb$_\mathrm{ss}$  & \textbf{1.17x} & \textbf{1.6690} & \textbf{1.2127} & \textbf{0.4564}
\end{tabular}
\end{table}

%% file: blockMerge_new.tex
\subsection{Block merges and segment remapping}
\label{sec:blockMergesec}
\begin{figure}
    \centering
    \includegraphics[width=0.5\textwidth]{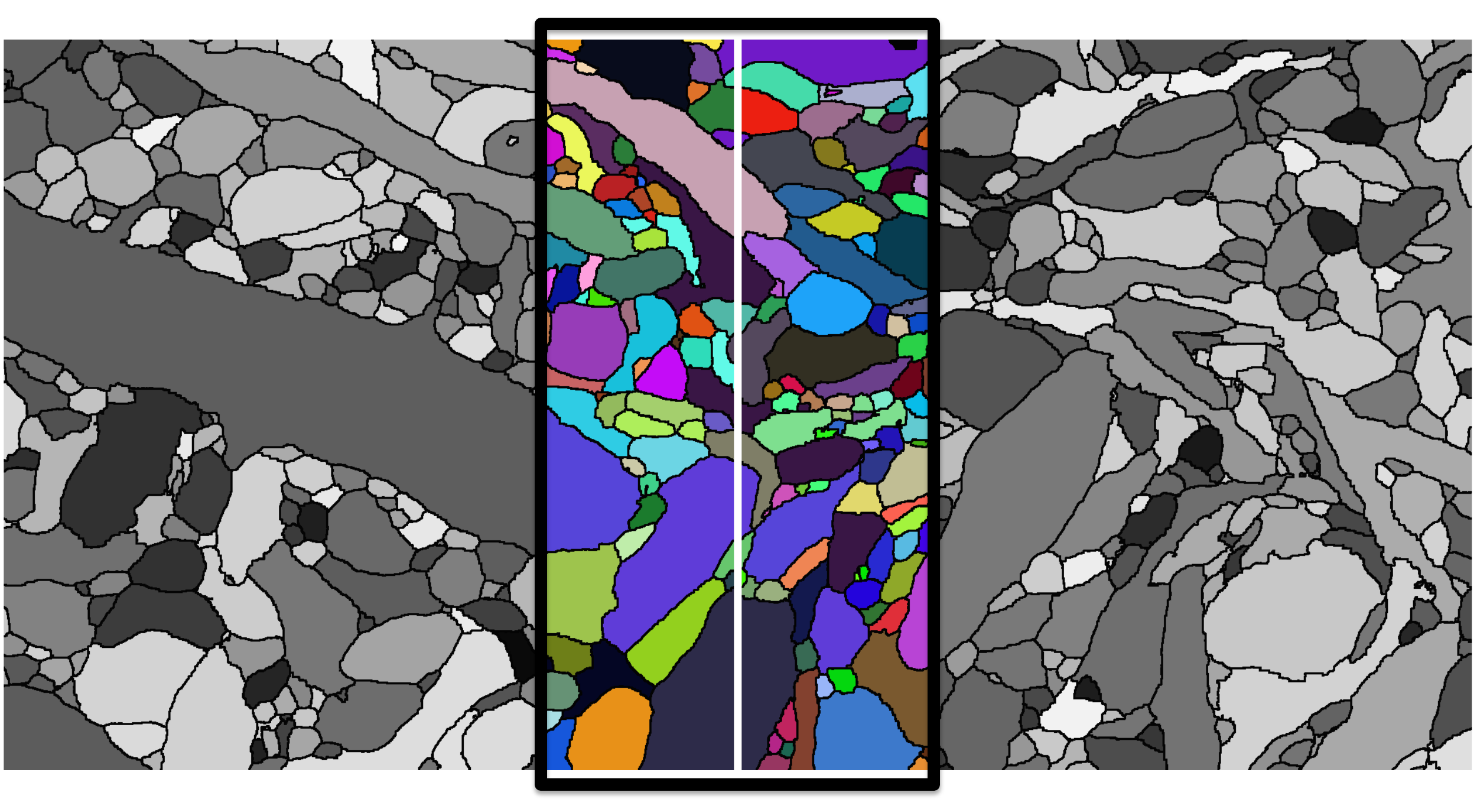}
    \caption{A 2D section of the volume showing two adjacent blocks and a block-slice, which is enclosed in a frame.}
    \label{fig:block_slice}
\end{figure}
We demonstrate an algorithm for matching neurites between two adjacent blocks by analyzing a thin volume around the shared face, which we term the \emph{block-slice} (see Figure \ref{fig:block_slice}). The merging algorithm pairs up corresponding neurite IDs, or \textit{merge-pairs}, across the two blocks.  Our two-step block-merging approach relies on the random-forest algorithm of Parag \emph{et al.}~\cite{parag2015context}. Both steps of our algorithm require EM images and predicted membrane probability maps from the block-slice.


First, we over-segment the block-slice using segmentations of the individual blocks, as shown in the highlighted area of Figure \ref{fig:block_slice}.  We agglomerate the over-segmentation using random-forests as implemented by NeuroProof \cite{matveev2016, parag2015context}. Specifically, NeuroProof outputs a set of merged IDs, of which we use only the ones crossing the boundary between the two blocks. 
In order to perform this step, we have trained a dedicated random-forest classifier. Note that the random-forest classifier used for the agglomeration within blocks would be inaccurate at predicting correct merges since the feature statistics of neurites within blocks are significantly different from those of neurites in block-slices: the latter are composed of objects within two sub-blocks, having distinct IDs. Below we refer to the segmentation resulting from this procedure as $S_1$.

The approach above improves accuracy (measured by VI, Equation (\ref{eq:vi})) with respect to merges of relatively small objects, as these merges govern the distribution to which the random-forest is exposed during training. However, our experiments show that often large objects, sometimes spanning the entire block-slice, are merged poorly. To overcome this issue, we re-segment the block-slice by computing over-segmentation and agglomeration using the in-block segmentation approach (with the same random-forest classifier). Importantly, this second-pass segmentation is not affected by any artificially defined boundary between the original blocks. We refer to this segmentation as $S$.

We next treat $S$ as a ground truth segmentation for the block-slice and compute the merge pairs in $S_1$ that improve the variation of information (equation) between the merged segmentation and $S$. For each candidate merge-pair we compute the change in the merge- and split-VI from applying a merge to $S_1$, respectively $\Delta_{merge}$ and $\Delta_{split}$. If $\Delta_{merge} / \Delta_{split} < K$, \emph{i.e.} by merging we improve the split VI without damaging the merge VI, then the merge is made, where $K$ is a bias parameter determined on validation set (set to $0.125$ for Kasthuri). Finally the merge-pairs from both steps are globally remapped across all block-slices, using a disjoint-set data structure.

To assess the quality of the merging technique, we have randomly sampled 30 volumes within our dataset that correspond to different pairs of adjacent blocks. We segmented each pair of adjacent blocks both using the block merging algorithm above and by the in-block agglomeration approach. The average VI metric between pairs of volumes segmented by these two approach was approximately 0.15, which means that the block merging technique closely matches the quality of in-block segmentation. Furthermore, heuristic block merging approaches are require hand-crafed parameter to deal with edge cases as well as sensitive to the specific algorithm used for in-block segmentation. By contrast, our block merging technique can transparently use any in-block segmentation algorithm as a backbone and hence benefits from any improvement made to it.

%% file: skeletonization_new.tex
\subsection{Skeletonization} 
\label{sec:skeletons}
\captionsetup[subfigure]{skip=3pt}
\begin{figure*}[t]
    \centering
     \begin{subfigure}[b]{0.24\textwidth}
        \includegraphics[scale=0.095]{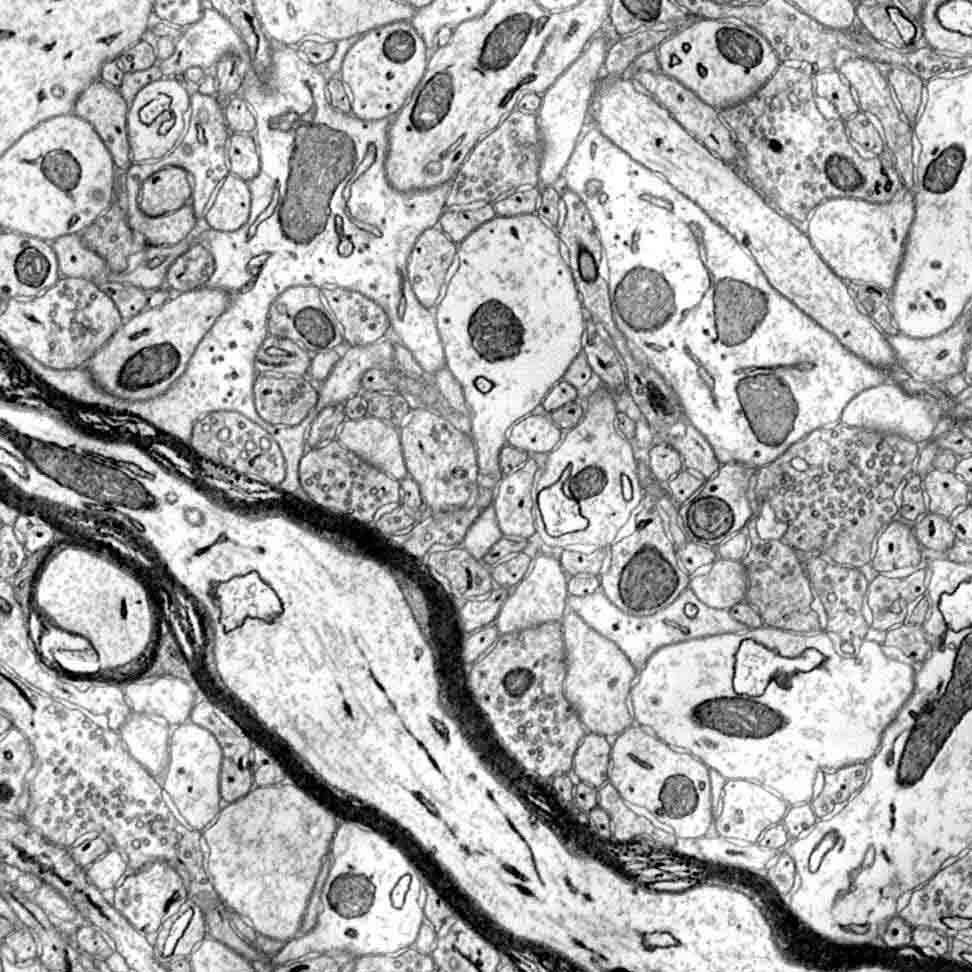}
        \caption{}
        \label{fig:skel_rec:gt}
    \end{subfigure}
    \hfill
    \begin{subfigure}[b]{0.24\textwidth}
        \includegraphics[scale=0.095]{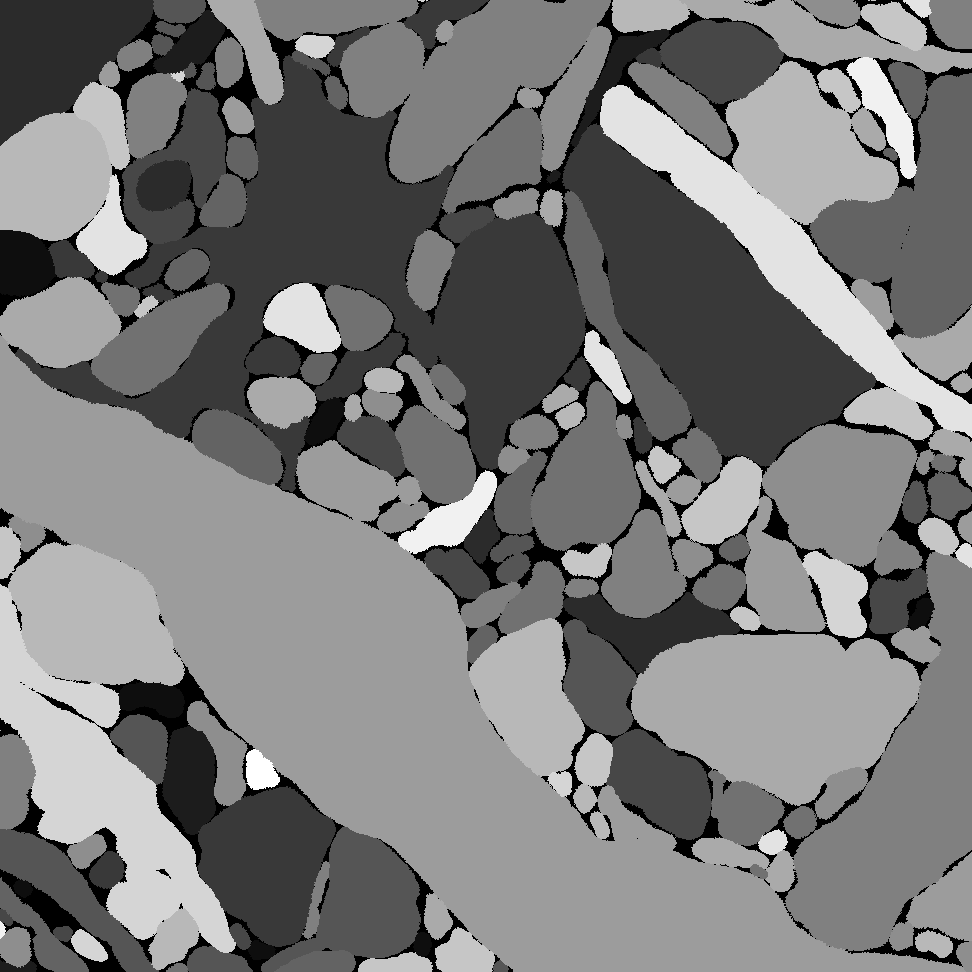}
        \caption{}
        \label{fig:skel_rec:gt}
    \end{subfigure}
    \hfill
    \begin{subfigure}[b]{0.24\textwidth}
        \includegraphics[scale=0.095]{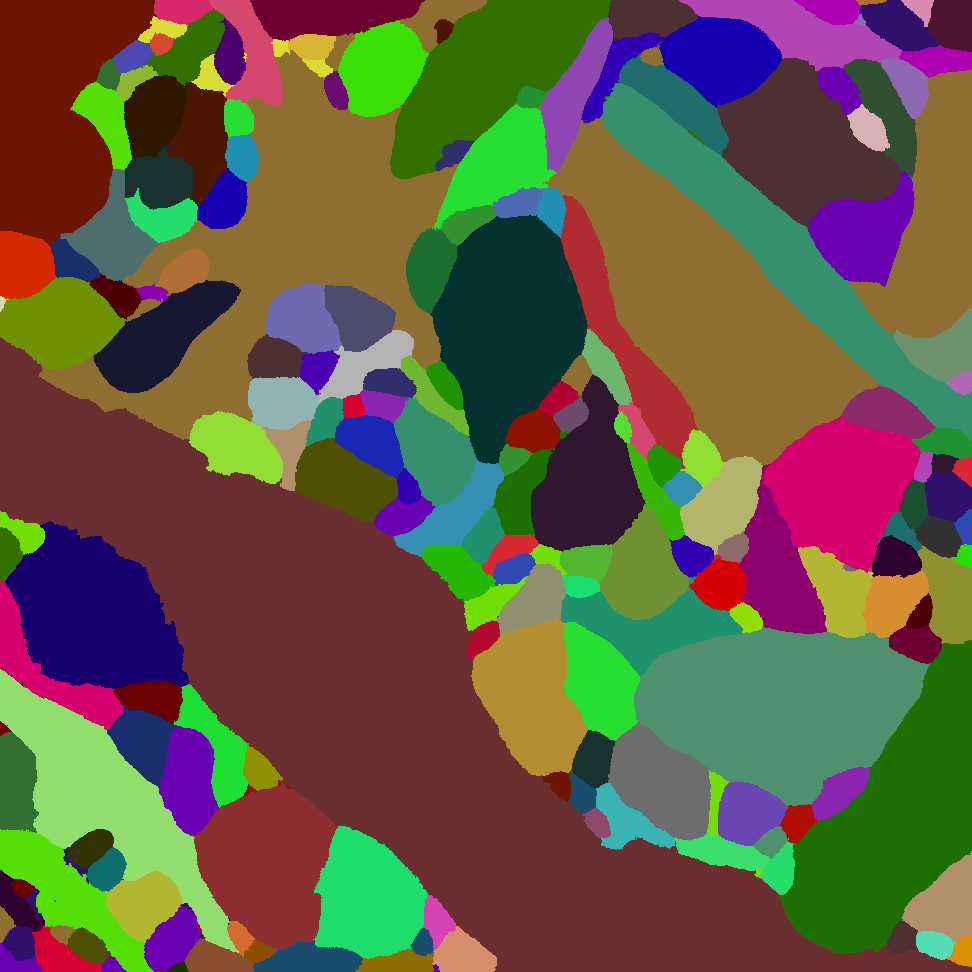}
        \caption{}
        \label{fig:skel_rec:seg}
    \end{subfigure}
    \hfill
    \begin{subfigure}[b]{0.24\textwidth}
        \includegraphics[scale=0.095]{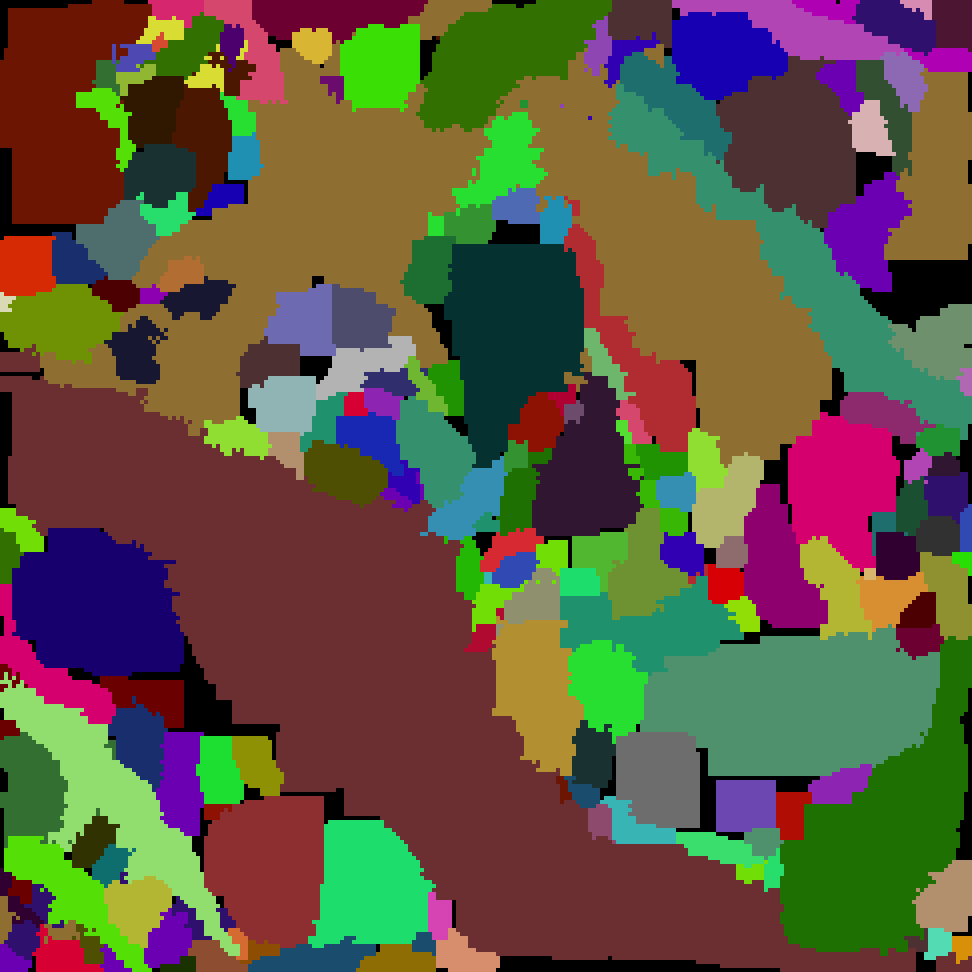}
        \caption{}
        \label{fig:skel_rec:rec}
    \end{subfigure}
    \caption{One slice of (a) EM, (b) ground truth, (c) segmentation, (d) skeleton reconstruction of AC3. The VI between the ground truth and reconstruction volumes was 3.0 (2.5 excluding boundaries).}
    \label{fig:skeleton-reconstruction}
\end{figure*}

To reconstruct neuron morphology from our segmentation, we perform 3-D skeletonization with a \emph{thinning} algorithm. Thinning algorithms delete voxels iteratively from an object until only a 2-D representation along the medial axis remains~\cite{saha_survey_2016}. At each iteration, only voxels on the boundary that do not change the topology of the object are deleted (so-called \emph{simple points}). Our simple point criterion is based on \cite{she_improved_2009}. The number of iterations in deleting voxels is recorded and saved as the skeleton \emph{width}. We parallelize the steps above using chromatic scheduling \cite{kaler_executing_2014}; as a result our implementation is equivalent to an 8-subfield thinning algorithm \cite{bertrand_three-dimensional_1995}.

A novelty of our approach is to perform the skeletonization on coarsened images. Our implementation of coarsening transforms each voxel into a larger voxel which contains a set of all labels inside. Besides reducing complexity while still preserving connectivity, our coarsening also allows us to correct for the anisotropic image resolution of the microscope (we coarsen by a factor of $(4c, 4c, c)$ in the $(x,y,z)$ directions where $c$ is the coarsening factor). After thinning, 3-D images of the segments are transformed into tree graphs based on 26-connectivity and are outputted into the .swc format \cite{feng_neutube_2015}.

\subsubsection*{Skeletonization accuracy and skeleton-volume expansions}
Besides preserving the topology of the segments, we propose that our skeletonization algorithm better captures the shape of the 3-D segments. Since there are no standard error metrics for skeletonization \cite{saha_survey_2016}, we propose to evaluate skeletons using the VI metric (used throughout for segmentation evaluation) after simulating a full-volume reconstruction using only the skeleton cross-sections (Figure \ref{fig:skeleton-reconstruction}).

Volume reconstruction from skeletons is performed in parallel using a thickening algorithm on an image coarsened by the same parameters as skeletonization. We grow balls of predetermined radius around each skeleton node, where the radius is computed by multiplying the skeleton width by a factor of $0.7$ (empirically tuned) that accounts for the speed of the thinning. In case the thickened balls intersect, we break collisions for each voxel based on the difference of ball radii and distance to the ball centers. When there is no unique highest difference, the label of the object is set to zero (background label).

The VI of our skeleton-based reconstruction with respect to ground truth is 3.03 (Figure \ref{fig:skeleton-reconstruction}). Similar to previous studies \cite{arganda2015crowdsourcing}, we also excluded ground truth boundary voxels from both image stacks and recalculated the VI as 2.52. This substantial improvement in our score indicates that our reconstruction successfully captures the high-level structure of the neurites and that much of the error is accounted for by regional boundaries.

%% file: errorDetection.tex
\section{Error detection and slow-pass reconstruction}
\subsection{Detecting merge errors}

Merge errors occur when two or more distinct objects are erroneously merged.  These represent an excellent test case for automatic reconstruction algorithms, since they are typically obvious to a human proof-reader, but are very challenging for an algorithm to identify due to the multiplicity of possible junctures between incorrectly merged objects~\cite{plaza2014focused}.

We outline an approach for detecting merge errors, using the following insight: Neurons rarely form an ``X'' shape in their branch structure.  However, X-junctions commonly occur when two distinct neurites pass close enough together that they are erroneously merged by automated segmentation algorithms. 
\begin{figure}
\centering
\includegraphics[width=1\linewidth]{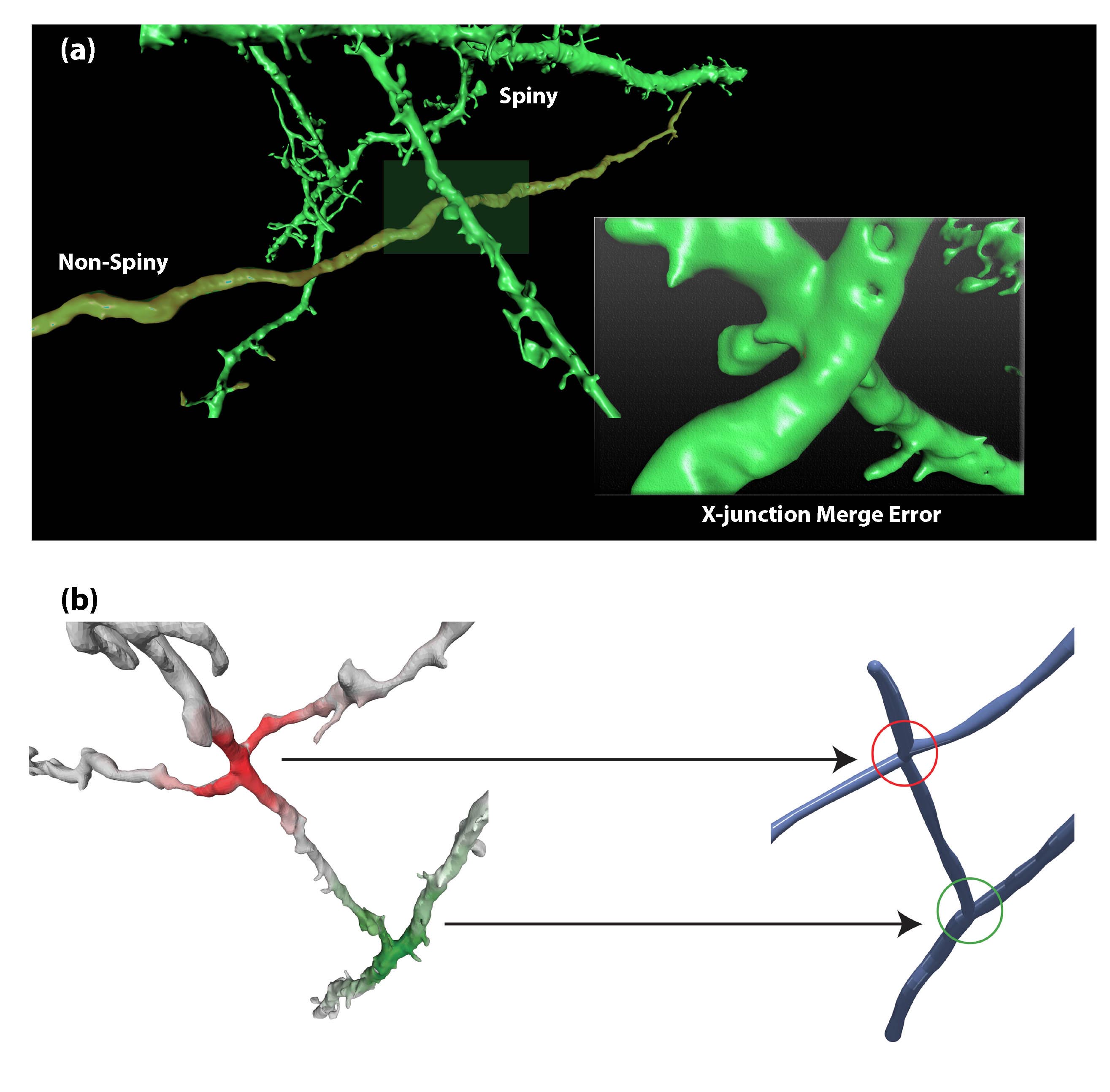}
\caption{(a) A merge error in the fast-pass pipeline resulting in the agglomeration of two processes with different morphology. In multi-pass approach this merge error can be identified based on geometry (as an X-junction), as well as morphology (one branch is a \emph{spiny} dendrite and the other is an axon), (b) an X-junction merge error and a normal branching process in a neuron, both correctly marked by our algorithm.}
\label{fig:crossRepair}
\end{figure}

To identify X-junctions within an object, we first construct a skeletal representation of the object as a tree (in the graph theoretical sense). Nodes in the tree correspond to branch points (satisfying degree $>$ 2), non-branch points that interpolate the curve along a single neuronal process (degree $=$ 2), and endpoints of branches (degree $=$ 1).  In order to simplify the representation, we eliminate degree-2 nodes, instead connecting their neighbors directly by edges within the graph.  This simplifies long processes that lack branch points.  Branches shorter than a specified threshold are then trimmed recursively, yielding a \emph{layout graph} that captures the essential structure of the neuron.

Next, the embedding of the layout graph in $\mathbb{R}^3$ is smoothed.  If the position of node $v$ is given by $\textbf{x}(v) \in \mathbb{R}^3$, then smoothing reassigns each position $\textbf{x}(v)$ simultaneously according to the positions $\textbf{x}(u)$ of neighbors $u\sim v$:
$$\textbf{x}(v) \gets \frac{\textbf{x}(v) + \frac{1}{d(v)}\sum_{u\sim v}\textbf{x}(u)}{2},$$
where $d(v)$ denotes the degree of $v$.  This smoothing allows the direction of neuronal processes to be estimated accurately without being sensitive to minor bends and noise.

Finally, forks in the smoothed layout graph are characterized as plausible or implausible.  The algorithm looks at every fork with four or more branches, or pairs of forks that are very close together and together have four or more branches.  If the branches can be paired such that direction and radius are approximately preserved across each pair, then the location is marked as implausible (see Figure \ref{fig:crossRepair}).

We evaluated our algorithm on the 100 largest objects in our segmentation, which are expected to have the highest incidence of merge errors (with the exception of the largest, which was overly noisy).  The algorithm correctly identified 47 X-junctions.  Three marked nodes arose from cell somata, and may be easily labeled as such, and one node was neither soma nor an X-junction.  Eight additional nodes were impossible to classify.  Not all X-junctions or merge errors are identified by this algorithm, and this ``false negative'' rate is harder to quantify without recourse to a larger dataset of ground truth; in either event, the error rate is substantially improved.

This algorithm for identifying X-junctions represents the first step in the correction of errors incurred within the standard segmentation pipeline. The locations identified as incorrect merges can be re-evaluated by the merge algorithm with a high threshold set for merging operations~\cite{plaza2014focused}.

\subsection{Extending segmentations}

Above we described a fast-pass pipeline for large-scale connectomics and an error-detection mechanism targeted at correcting morphological errors. We now present a slow-pass algorithm for extending a correct segmentation in order to correct errors or confirm uncertain regions of interest. Our algorithm described herein employs represents an extension of our fast-pass CNN architecture (FastMemb), with additional layers and features. The algorithm recursively increases the volume in which an object segmentation is correct by using the object geometry, membrane probability map and electron microscopy images in order to predict the object masks adjacent to the volume's boundary. We demonstrate repair of neurites wrongly segmented by the fast-pass pipeline by isolating and extending sections in which segmentation is correct. 

\subsubsection*{MaskExtend: sparse segmentation via CNN}
\begin{figure}
\centering
\includegraphics[width=1\linewidth]{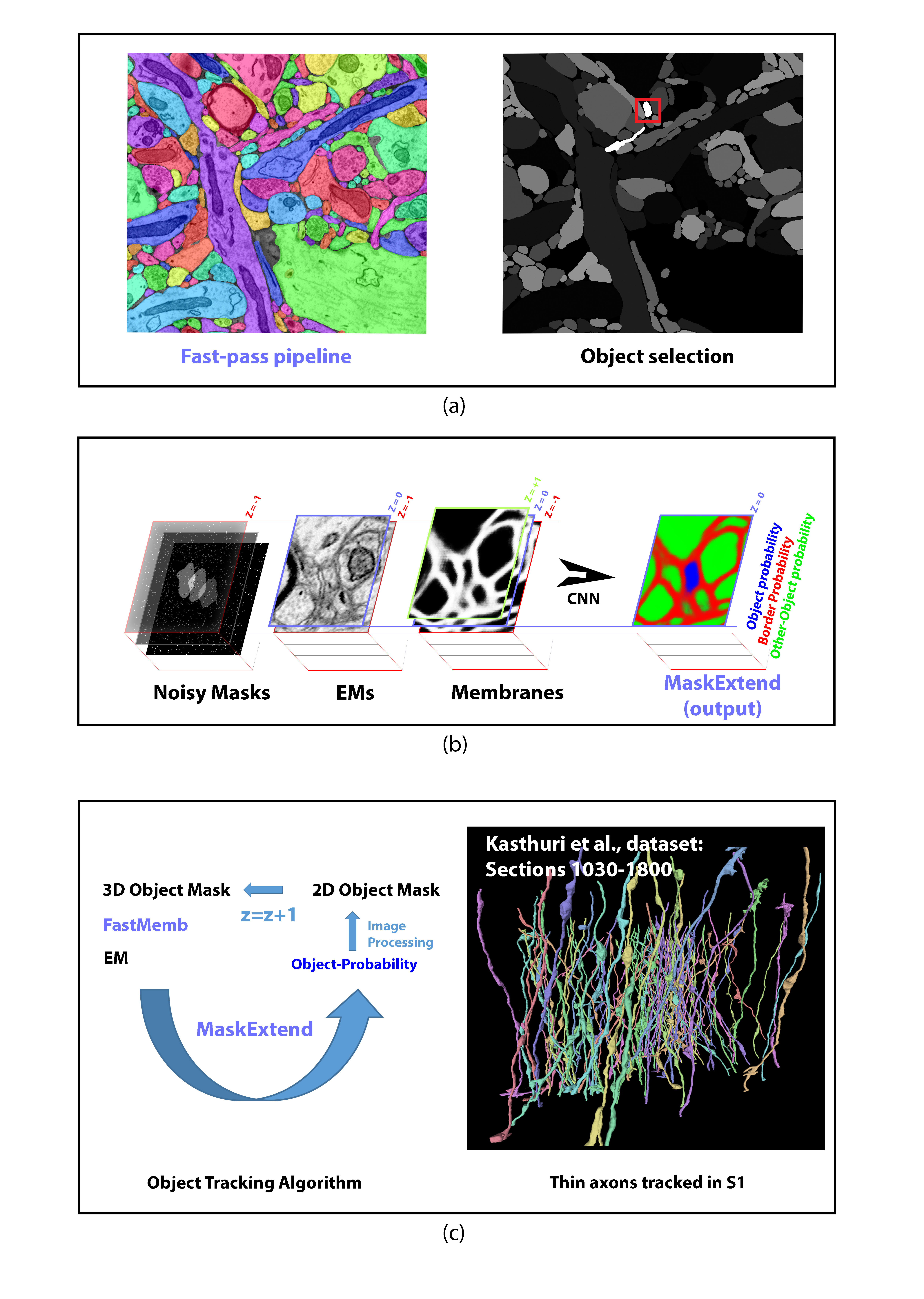}
\vskip -0.4in
\caption{MaskExtend slow-pass sparse segmentation. (a) Fast-pass input to sparse segmentation, (b) CNN architecture of MaskExtend, (c) axon morphology of all neurites crossing section 1031 in Kasthuri (${\sim}1031\times 30$nm vertically to the base of the cropped cortical tissue) and reaching section 1800 ({\thickmuskip=0mu${\sim}1800\times 30$} nm from baseline). To demonstrate the ability to reconstruct thin neurites, only thin intersections (cross section area smaller than 0.022{\textmu}m$^2$) were extended.}
\label{fig:maskextend}
\end{figure}
The computational approach we have thus far presented employs a minimalistic design; we strived to minimize runtime of all computational units, adhering to the microscope's pace, while keeping reconstruction quality close to state of the art (see also \cite{matveev2016}) and presented a mechanism for detecting morphological errors produced by the fast-pass pipeline. Below we describe how corrections of individual segments may be carried out with focused computational effort.

Recently \cite{januszewski2016flood} presented a CNN that expands predetermined seed locations in EM volumes while respecting neuronal boundaries, and iteratively reconstructing complete neuronal processes. Their approach shows for the first time that neural nets can be trained to output directly a highly accurate (sparse) segmentation. In parallel and independently of their work, we here present a CNN that extends an object mask from a small volume to a larger one. We posit that extending a single object mask is easier than classifying membrane voxels, since a partially revealed geometry of the object can impose a strong prior on the global shape of the object.

The CNN architecture used here for sparse segmentation, termed \emph{MaskExtend}, extends the max-out architecture of FastMemb, both in depth (+1) and number of features per layer. Figure \ref{fig:maskextend}(b) demonstrates the operation of MaskExtend on a 3-D patch comprised of several inputs. MaskExtend operates on a stack of aligned input images consisting of three consecutive EM images ({\thickmuskip=0mu$z=-1,0,+1$}), four consecutive binary masks of an object ({\thickmuskip=0mu$z=-4,\dots,-1$}) and five consecutive membrane probability maps ({\thickmuskip=0mu$z=-2,...,+2$}), collectively constituting twelve 2-D feature maps.

The network classifies the center pixel of the EM input (at {\thickmuskip=0mu$z=0$}) based on whether it belongs to the object represented by the mask images (at {\thickmuskip=0mu$z=-4,\dots,-1$}).  Thus, one output channel marks pixels of the desired neuron and (with some useful redundancy) a second channel marks pixels from other neurons. We dedicated a third output channel to mark membranes, which has the potential to locally and recursively improve training data for the fast-pass network FastMemb. Inference is computed in fully-convolutional mode to produce a new object probability image. In the next step, the object probability is smoothed and thresholded to form a binary mask of the predicted object, augmenting the object mask used as input to the network. The algorithm continues iteratively in either the $z+$ or $z-$ direction. For extending in $z+$, the algorithm computes the object's mask at section $z$, \emph{Mask$_z$}, recursively,
\begin{align}
\label{eq:transfer}
 \text{Mask}_z = J(\text{CNN}&(\textnormal{Mask}_{z-3,...,z-1}, \\  \nonumber
 			&\text{EM}_{z-1,...,z+1}, \\ \nonumber
 			&\text{FastMemb}_{z-1,...,z+1})),  
 \end{align}
where CNN stands for the neural network's dense inference, EM is the electron microscopy image stack, FastMemb is a stack of membrane probabilities and $J$ is a spatial smoothing and pixel thresholding operation applied to the output of the neural network.

MaskExtend's architecture was designed to optimize for accuracy instead of speed, as compared to the fast-pass pipeline. MaskExtend is comprised of four identical layers of alternating convolution/max-pool pairs (+1 depth with respect to fast-pass) aggregated with a max-out function \cite{goodfellow2013maxout} and an additional convolution. Identically to FastMemb, convolution layers of MaskExtend use stride-1 {\thickmuskip=0mu$4\times 4$} kernels, whereas number of features is increased from one layer to the next (32, 40, 48, 56 and 64, for five convolutions respectively). Combined with stride {\thickmuskip=0mu$2\times 2$} fully-convolutional max-pooling this architecture yields a {\thickmuskip=0mu$109\times 109$} field of view (FoV) which is roughly twice as large as that of FastMemb.

\subsubsection*{Training}
\begin{figure}
\centering
\includegraphics[width=1\linewidth]{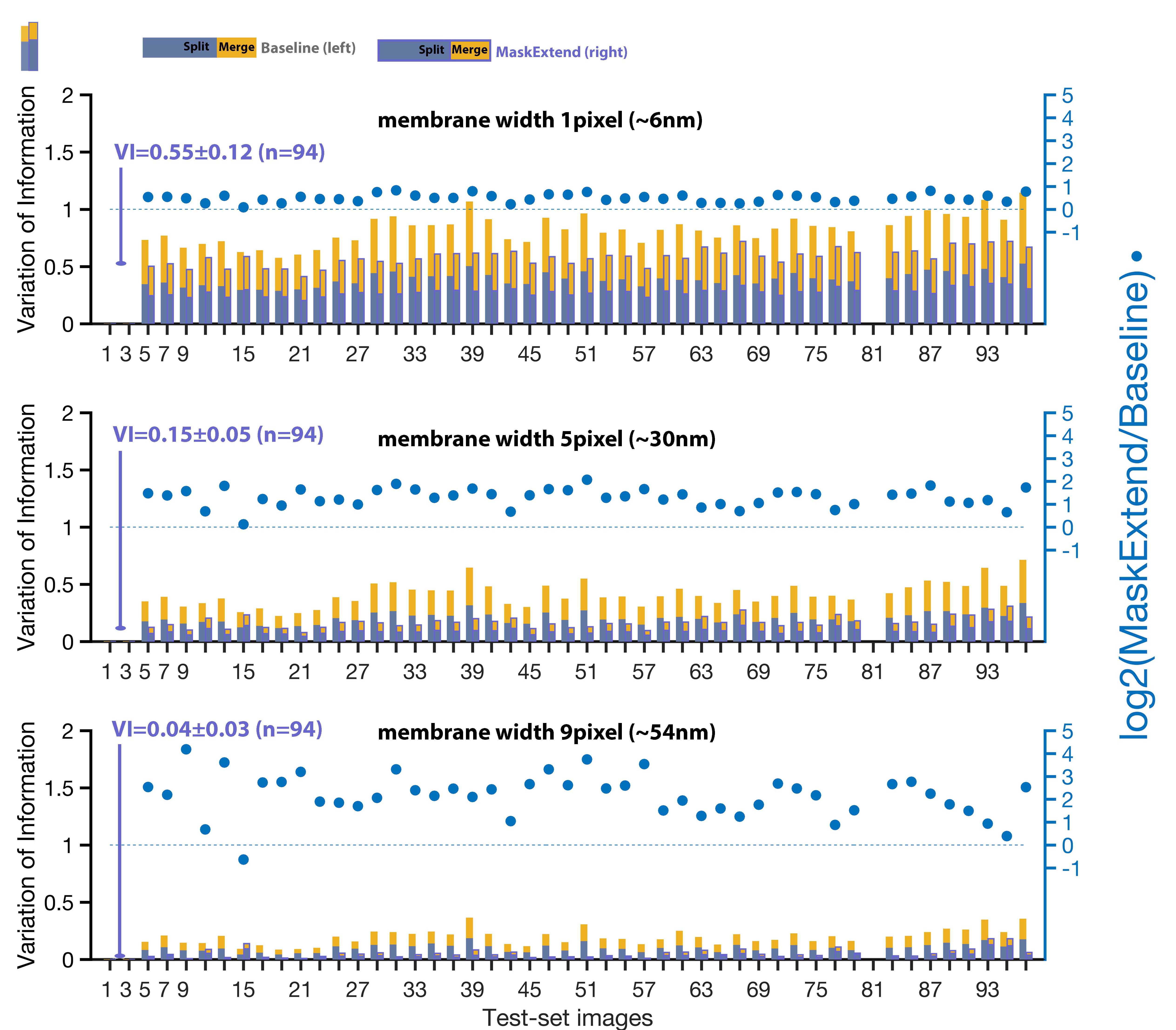}
\caption{Evaluating the transfer function of MaskExtend (Equation \ref{eq:transfer}). The plots display the varition of information~(VI; Equation \ref{eq:vi}) of single-section predictions using MaskExtend and a baseline prediction. The baseline prediction was defined by keeping the segmentation unchanged from one section to the next, representing the VI change in the ground truth segmentation. Three membrane thicknesses were imposed on the segmentation and discarded from evaluation as in previous studies. Right axes present the VI log-ratio improvement of MaskExtend over baseline. The largest improvement achieved for thick membranes suggests that MaskExtend is adequate as a correction algorithm for split errors that are incident to specific erroneously segmented sections, while producing few merge errors (negligible VI). Figure \ref{fig:maskextend}(c) shows the large-scale result of applying MaskExtend to cross sections of axons in S1~\cite{kasthuri2015saturated}.}
\label{fig:maskextend-VI-metrics}
\end{figure}
Training procedures and hyper-parameter selection are similar to those used in training FastMemb with few exceptions.  MaskExtend extends object masks forward and backward along a stack of EM images; however, we are limited in our ability to benchmark the accuracy of such an algorithm.  Our ground truth datasets (AC3 and AC4) include approximately 500 million annotated pixels of several hundred objects (dense annotation) within a small volume of less than 12 {\textmu}m$^3$.  We have therefore deferred 3-D morphology accuracy measurements until a benchmark for single-neuron reconstruction is created. The extension algorithm described above calls the MaskExtend CNN recursively. Hence, is it essential to evaluate the performance of the transfer function that computes a single mask based on preceding masks and fast-pass input (leading to the axon reconstruction in Figure\ref{fig:maskextend}(c)). We predicted neuron masks in all manually annotated sections in AC4 ({\thickmuskip=0mu$z=4,\dots,98$}), based on the diagram in Figure \ref{fig:maskextend}(b), and compared in Figure \ref{fig:maskextend-VI-metrics} the variation of information between the predicted 2-D segmentation and manual annotation. For brevity we display the accuracy of predicted object mask for twenty-eight predictions based on ground truth segmentation. As a baseline we also display the VI of the baseline prediction that predicts no change in segmentation from one slice to the other, which is an entropy measurement of the change in information between two adjacent ``segmentation truths.'' Evaluations were made for the non-membrane objects by imposing varied membrane width on the segmentation as in previous studies (panels in Figure \ref{fig:maskextend-VI-metrics}). Since variation between consecutive images reduces when neglecting information near object boundaries, we have also measured the log-ratio between the accuracy of MaskExtend, relative to baseline (Figure \ref{fig:maskextend-VI-metrics}, right axes). The improvement of MaskExtend over baseline is apparent in all membrane definitions and is significantly largest when considering thick membranes. This suggests that merge errors are scarce and that erroneous splits of 2-D objects appearing in difficult EM sections can be corrected based on fairly well-segmented neurons in adjacent sections. Below we demonstrate an extension of hundreds of axons of a single segmentation section computed by fast-pass upward ($z+$) to the entire S1 volume (see Figure \ref{fig:maskextend}(c)). 

MaskExtend was trained with the AC3-256 manually annotated dataset \cite{kasthuri2015saturated} in the following way. We used one million examples for each of the three categories as explained above: Mask, Non-Mask and Membrane. We used extracellular space annotations as an independent object different from the Membrane category which was processed as in FastMemb. For the Mask and Non-Mask categories we randomized centers of volumetric patches and used each to generate both a Mask and a Non-Mask category. Mask examples were processed by setting to zero pixels of objects in the mask input that correspond to different ground truth labels. Non-Mask examples were processed by randomly picking a random label from the ground truth labels based on the masked input and setting all other mask pixels to zero. This novel training procedure allows the neural net access to a larger pool of training examples. In addition, we applied binomial noise to all masks by randomly flipping mask pixels (see Figure \ref{fig:maskextend}(b)). Training was halted after convergence at 100K iterations, with accuracy on validation set nearly 95\%.

\subsubsection*{Object tracking}
Figure \ref{fig:maskextend}(c) demonstrates the results of MaskExtend applied to a set of neuronal sections identified by the fast-pass pipeline. Initial object masks were selected based on two criteria: (a) the area of the object should be small since thin objects tend to develop broken morphology and (b) the neuron IDs should be spatially consistent in preceding sections. We run this sparse segmentation on all cross sections incident to section {\thickmuskip=0mu$z=1031$} in the Kasthuri dataset and depicted in Figure \ref{fig:maskextend}(c) all neuronal processes that reached the dataset ceiling at {\thickmuskip=0mu$z=1800$}. The algorithm reconstructed more than hundreds of axons, hundreds of axon boutons, including \emph{en passant} boutons that are clearly visible in \ref{fig:maskextend}(c).

%% file: conclusions_new.tex
\section{Conclusions}
\label{sec:reconstruction}

We present a time-constrained approach to the big data challenge of connectomics, complementing previous systems that optimize for accuracy alone \cite{knowles2016rhoananet,matveev2016}. Our fast-pass pipeline (CNN architecture, optimized watershed and multi-core agglomeration) attains microscope pace without considerable loss of accuracy compared to previous solutions, and outperforms the accuracy of previous pipelines benchmarked for speed \cite{PlazaBerg2016,roncal2015automated}. The speed of the fast-pass algorithm allows us to introduce a slow-pass framework for error detection and correction.  We present slow-pass algorithms for morphological X-junction detection and for spatially extending segmentation using a novel CNN architecture.

To demonstrate the speed and accuracy of our fast-pass pipeline, we performed two full reconstructions of the Kasthuri dataset and each of the manually annotated test sets (AC3-256, AC3-75), one with initial images at 3nm pixel resolution (yielding the largest automatic reconstruction of a cortex EM dataset) and one at 6nm pixel resolution as used in the ground truth labeling. Our 6 nm execution took 5 hours on the entire S1 dataset and corresponded to a VI of 1.83 on the AC3-256 test set (trained on AC4-100), while our subsampled implementation over 3 nm data improved this VI to 1.66 (details of our multi-core implementation appear elsewhere \cite{matveev2016}). The previous-best results benchmarked for speed of a fully automated connectomics pipeline took 3 weeks to execute on a farm of GPUs and CPUs \cite{kaynig2015large} and reported a less accurate VI of 1.99, compared to 1.77 when evaluating our pipeline on the same portion of the test set \cite{kaynig2015large}. (See also the recently revised RhoANA pipeline published by several of the authors \cite{knowles2016rhoananet}.)

\begin{figure}[t]
\centering
\includegraphics[width=1.0\linewidth]{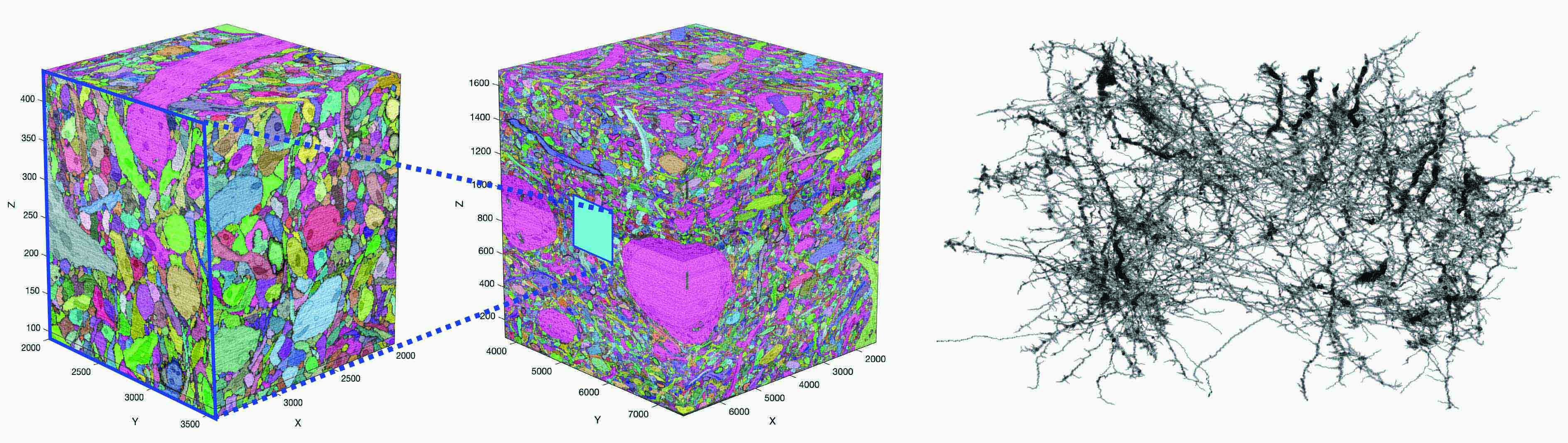}
\caption{3-dimensional segmentation (left) and skeletonization (right) of the full Kasthuri dataset ({\thickmuskip=0mu$\sim$100,000} {\textmu}m$^3$ of cortex~\cite{kasthuri2015saturated}, representing 473 GB). The leftmost segmentation shows a small-scale automatic reconstruction ({\thickmuskip=0mu$\sim$850} {\textmu}m$^3$) compared to previous large-scale reconstruction attempts ($\sim$27,000 {\textmu}m$^3$ \cite{kaynig2015large}. The segmentation on the right shows what is possible with our high throughput pipeline (order of {\thickmuskip=0mu$\sim$100,000} {\textmu}m$^3$ per day).}
\label{fig:neuronblock}
\end{figure}

In addition, the pipeline presented here can compute segmentation on the full Kasthuri dataset ({\thickmuskip=0mu$\sim$100,000} {\textmu}m$^3$) (Figure~\ref{fig:neuronblock}), in only 5 hours on a single multicore machine \cite{matveev2016}, compared to an extrapolated runtime of three weeks for previous pipelines running on a farm of GPUs and CPUs \cite{kaynig2015large,PlazaBerg2016}. To our knowledge this is the first automated reconstruction of the full S1 volume.


